\documentclass[10pt,twocolumn]{revtex4-1} 
\usepackage{bm}
\usepackage{hyperref}
\usepackage{makeidx}
\usepackage{amsmath}
\usepackage{dcolumn}
\usepackage{comment}
\usepackage{color}
\usepackage{amsfonts}
\usepackage{amssymb}
\usepackage{epsfig}
\usepackage{float}
\usepackage{cleveref}
\usepackage[british,UKenglish,USenglish,english,american]{babel}
\usepackage{epstopdf}
\usepackage{array}
\usepackage{physics}
\usepackage{graphics}
\usepackage{subcaption}

\usepackage[justification=justified,singlelinecheck=false,font=small]{caption}
\usepackage{subcaption}
\usepackage{ragged2e}

\makeatletter
\long\def\@makecaption#1#2{%
  \vskip\abovecaptionskip
  \begingroup
    \small
    \parindent=0pt
    \parbox{\columnwidth}{\justifying\textbf{#1.} #2}%
  \endgroup
  \vskip\belowcaptionskip
}
\makeatother

\begin{document}

\title{ Domain-Wall Control of Topological Qubits in the Kitaev–SSH Chain}


\author{M.~A.~R.~\surname{Griffith}$^{1}$}
\email{griffithphys@gmail.com}

\author{S.~\surname{Rufo}$^{2,3}$}
\author{H.~\surname{Caldas}$^{4}$}
\author{Rosiane de Freitas$^{2}$}

\affiliation{$^{1}$Departamento de Física, Universidade Federal do Amazonas, 
Av.\ Gen.\ Rodrigo Octávio 6200, Coroado~I, 69080-900 Manaus, Brazil}

\affiliation{$^{2}$Instituto de computação, Universidade Federal do Amazonas, 
Av.\ Gen.\ Rodrigo Octávio 6200, Coroado~I, 69080-900 Manaus, Brazil}

\affiliation{$^{3}$Instituto Militar de Engenharia IME, 
Praça Gen. Tibúrcio, 80 - Urca, Rio de Janeiro - RJ, 22290-270, Brazil}

\affiliation{$^{4}$Departamento de Ciências Naturais, Universidade Federal de São João del Rei,
Praça Dom Helvécio 74, 36301-160 Sao João del Rei, MG, Brazil}

\date{\today }

\begin{abstract}
Zero-energy states in one-dimensional SSH–Kitaev hybrid systems have emerged as promising candidates for topological qubits. In our work, we show that introducing a domain wall into a chain with anisotropic superconducting correlations provides a powerful way to control both the number and the nature of these boundary modes. The defect acts as a digital knob: its presence or absence flips the parity of zero modes and thus decides whether an isolated Majorana exists at the chain ends. This “on/off” mechanism is significantly more robust and simpler than fine-tuning global parameters such as chemical potential or hopping amplitudes. Moreover, anisotropy provides an additional lever to calibrate the effect of the defect, opening a pathway to architectures where topological qubits can be locally addressed by domain walls. This proposal reframes defects not as imperfections, but as useful resources for quantum information and computation.
\end{abstract}

\maketitle

\section{Introduction}

One of the central challenges in quantum computing is protecting information against noise. Among the strategies proposed, topological qubits stand out because their stability does not rely solely on increasingly demanding experimental isolation, but on the inherent robustness of quantum states protected by symmetries and topology. In this setting, Majorana states in one-dimensional systems are particularly appealing, since they appear as boundary excitations that cannot be easily removed by local perturbations\cite{kitaev2001,Alicea_2012,Aguado2017,RACHEL20251,AliceaSpinHall}.

Kitaev-like models provide a rich platform, both theoretically\cite{kitaev2001, Alicea_2012, GriHeronPhysRevB} and experimentally\cite{SarmaRevModPhys,DasSarma2012},
for understanding how Majorana zero-energy bound states can be created and stabilized. Coupling to realistic substrates or environments inevitably introduces
anisotropies and noise that tend to suppress the topological phase and destabilize
the zero-energy bound states, including the Majorana quasiparticles\cite{GriHeronPhysRevB}.
\begin{figure}[t!]
    \centering
    \includegraphics[width=0.9\linewidth]{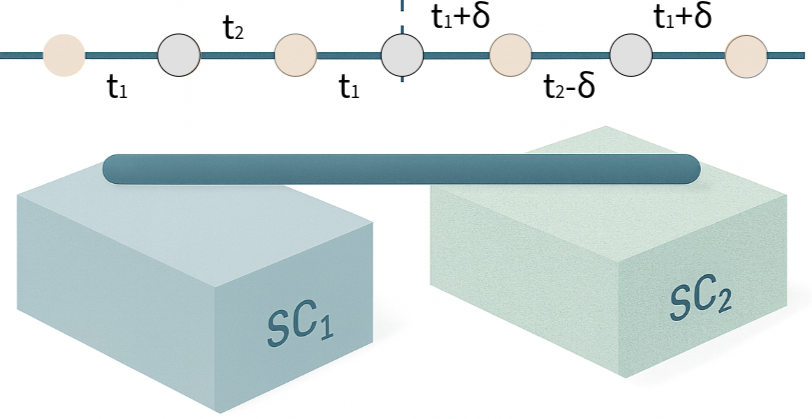}
\caption{Schematic representation of a spin--orbit--coupled nanowire deposited on top of two $s$-wave superconductors, for example, Al and Nb. An external magnetic field is applied along the wire, parallel to the superconducting surfaces.}
    \label{fig:placeholder}
\end{figure}
Recently, numerous theoretical and experimental works have proposed models and
strategies to mitigate the effects of defects and interactions that introduce
anisotropies or noise\cite{Capizzi_2023}. Defects, disorder, and environmental noise have profound effects on Majorana
zero modes on realistic platforms. Spatial inhomogeneities in the chemical
potential or pairing potential tend to localize or hybridize edge modes,
lifting the near-zero-energy degeneracy that protects the topological qubit
\cite{Lutchyn2010,Capizzi_2023,Alicea_2012}. Charge noise and fluctuations in gate
potentials produce random shifts in the effective mass term, which can push the
system through local topological--trivial boundaries and thereby destabilize the
Majorana pair \cite{DasSarma2012,Stanescu2013}. Coupling to phonons, residual
quasiparticles, or substrate excitations generates decoherence channels that
break fermion-parity conservation, a key requirement for topological protection
\cite{Rainis2012,Higginbotham2015}. Magnetic inhomogeneities and spin--orbit
disorder further distort the Bogoliubov--de Gennes spectrum, reducing the
mini-gap and enhancing the susceptibility to quasiparticle poisoning
\cite{Aguado2017}. These effects make the control of disorder and noise--- either
through material optimization or topological engineering strategies---essential
for stabilizing Majorana-based quantum devices.

In our work, we investigated a hybrid SSH–Kitaev chain with anisotropic superconducting correlations. This system reveals a striking feature: the introduction of a domain wall i.e., a topological defect, at the center of the chain fundamentally changes the counting of zero-energy states. Depending on the anisotropy, the defect generates a solitonic zero mode that modifies the boundary parity and, in doing so, determines whether an isolated Majorana exists at the edges.
This observation has a direct consequence for quantum computing: the defect functions as a digital switch. Without it, isolated Majoranas do not emerge; with it, the qubit becomes active. This contrasts sharply with the usual approach of tuning global parameters such as chemical potential or hopping amplitudes—an experimentally delicate and often unrealistic task. Instead, a single local defect is enough to turn on or off a qubit.
The key message is simple: defects, often regarded as imperfections, can be reinterpreted as resources. In this perspective, the domain wall is not a flaw to be eliminated but a control mechanism to be harnessed—a quantum switch to activate or deactivate topological information.

\vspace{1cm}
\section{Model}

We defined the Defected Kitaev–SSH Chain (DKSC) model as %
\begin{align}
\mathcal H
&= \sum_{i=1}^{N/2}
\Big[ \Big(t_1\, c^\dagger_{B i} c_{A i} + t_2\, c^\dagger_{A,\,i+1} c_{B i}  
+ \Delta_1\, c^\dagger_{B i} c^\dagger_{A i} \nonumber\\ 
&+ \Delta_2 c^\dagger_{A,\,i+1} c^\dagger_{Bi}+\text{H.c} \Big)
- \mu  \Big(c^\dagger_{A i} c_{Ai}+c^\dagger_{Bi} c_{Bi} \Big) \Big] \nonumber\\
&+ \sum_{i=N/2+1}^{N}
\Big\{ \Big[(t_1+\delta)\, c^\dagger_{B i} c_{A i} + (t_2-\delta)\, c^\dagger_{A,\,i+1} c_{B i}
\nonumber\\ 
&+ \Delta_1\, c^\dagger_{B i} c^\dagger_{A i} 
 + \Delta_2 c^\dagger_{A,\,i+1} c^\dagger_{Bi}+\text{H.c} \Big ] \nonumber \\
&- \mu  (c^\dagger_{A i} c_{Ai}+c^\dagger_{Bi} c_{Bi}) \Big\} .
\label{eq:realspace}
\end{align}
where, $c^{\dagger}_{\alpha,i}$ creates a fermion at site i and sublattice $\alpha=A,B$. $\mu$ is the chemical potential, $t_1$ and $t_2$ are the hopping terms, $\delta$ characterizes the distribution difference between the two half of the chain and $\Delta_i$ ($i=1,2$) are the anisotropic correlations of the superconduction spinless phase.  The Defected Kitaev–SSH Chain (DKSC) model describes a dimerized chain in the non-conventional p-wave or effective spinless  superconducting phase.

The SSH-like defect is introduced by making $t_1$ and $t_2$ different in the two halves of the chain (for example, $t_1\to t_1+\delta$ and $t_2\to t_2-\delta$ for the second half). A domain-wall can be generated in the Kitaev-SSH hybrid chain at any site of the chain. Here, for simplicity, we choose to create a topological defect at the middle of the chain. Therefore, the domain wall is obtained by breaking the dimerization at the site $N/2$ of the chain. This dimerization pattern can localize zero-modes at the interface, whose properties will be affected by the pairing amplitudes $\Delta_{1,2}$. 
The Kitaev-SSH hybrid chain is depicted in fig.~\ref{fig:placeholder}. In realistic devices, left–right asymmetries arise naturally when a single proximitized chain simultaneously overlaps with two distinct parent superconductors or substrates, leading to spatially dependent pairing and hopping amplitudes. Such setups provide a material route to engineer domain
walls in Kitaev--SSH–type hybrids, where the left and right halves of the chain
realize different effective parameter sets and host interface Majorana modes.

\paragraph{Localized zero energy states:}

We performed a real-space diagonalization of the Hamiltonian matrix for various unit-cell sizes $N$. 
The eigenvalues encode the collective excitation energies of the fermionic modes, while the corresponding eigenvectors are used to compute the probability density $\rho_i=|\Psi_i|^2$ at each site $i$where $\Psi_i$ are the Hamiltonian eigenvectors corresponding to the zero-energy modes. 

\section{Results and Conclusions}

\begin{figure}[t!]
    \centering
    \begin{subfigure}{0.23\textwidth}
        \centering
        \includegraphics[width=\textwidth]{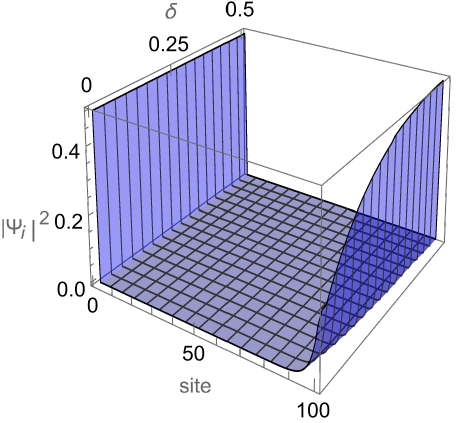}
        \caption*{(a)}\label{fig2a}
    \end{subfigure}
    \hfill
    \begin{subfigure}{0.23\textwidth}
        \centering
        \includegraphics[width=\textwidth]{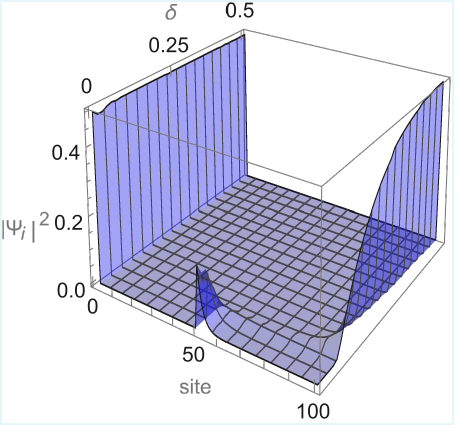}
        \caption*{(b)}
    \end{subfigure}

    \begin{subfigure}{0.23\textwidth}
        \centering
        \includegraphics[width=\textwidth]{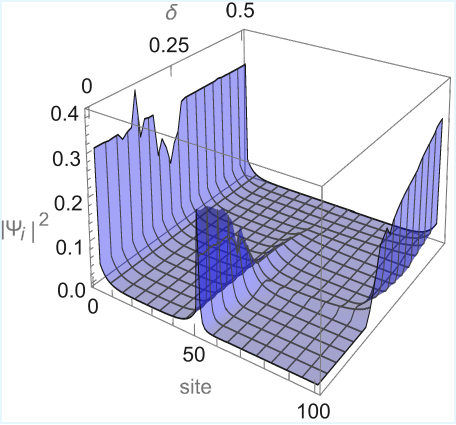}
        \caption*{(c)}
    \end{subfigure}
    \hfill
    \begin{subfigure}{0.23\textwidth}
        \centering
        \includegraphics[width=\textwidth]{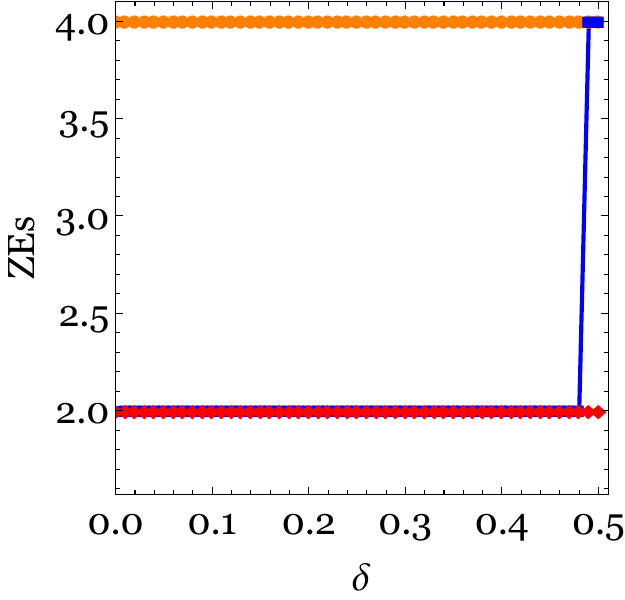}
        \caption*{(d)}
    \end{subfigure}
\caption{Probability density $\rho_i = |\Psi_i|^2$ at site $i$ as a function of $\delta$ for $\mu = 0$. 
Results are shown for different values of the pairing ratio $\Delta_2/\Delta_1$:  (a) $\Delta_2/\Delta_1 = 0.4$, (b) $\Delta_2/\Delta_1 = 0.6$, (c) $\Delta_2/\Delta_1 = 0.8$, and (d) Number of zero energy states as a function of $\delta$ }\label{figmuzero}
\end{figure}

\paragraph{Fine-tuning limit:} Figure \ref{figmuzero} shows the $\mu=0$ limit for $t_{2}/t_{1}=1.5$ and several values of $\Delta_{2}/\Delta_{1}$. For $\Delta_{2}/\Delta_{1}=0.4$ [Fig.~\ref{figmuzero}(a)], zero-energy modes emerge at the two ends of the chain. These non-local modes are the characteristic Majorana bound states, charge-neutral quasiparticles that are equal to their own antiparticles. 
The zero-energy mode localized at the left edge corresponds to the $\gamma_L$
Majorana, while the one at the right edge corresponds to $\gamma_R$.

For $\Delta_{2}/\Delta_{1}=0.4$ and $0 \le \delta \le 0.5$, a single Majorana bound
state emerges at each end of the DKSC. In this regime, these nonlocal modes are 
insensitive to the defect located at the center of the chain. However, when $\Delta_{2}/\Delta_{1} \ge 0.6$, the anisotropic pairing term becomes
dominant and a domain wall forms at the center of the chain. 

We find that for $\Delta_{2}/\Delta_{1}=0.6$ and small $\delta$, the zero-energy mode originally localized at the right edge progressively shifts toward the center of the chain, as shown in Fig.~\ref{figmuzero} (a)-(c). The total number of zero-energy modes remains equal to two, as illustrated in Fig.~\ref{figmuzero}(d) (blue curve). Fig.~\ref{figmuzero}(d) shows the number of zero-energy states as a function of $\delta$ for three different values of ratio $\Delta_2/\Delta_1=0.4 \text{(red)},0.6\text{(blue)},0.8\text{(orange)}$. Note that for $\Delta_{2}/\Delta_{1}=0.6$ the number of zero-energy states changes
from two to four. This behavior reflects the formation of a solitonic configuration: the domain wall nucleates a bound Majorana mode that absorbs the right-edge state and relocates it to the middle of the chain. Because the left-edge Majorana remains in place, the mid-chain solitonic mode can likewise be interpreted as a Majorana zero-energy bound state, preserving the nontrivial topology while modifying the spatial distribution of the zero modes. As $\delta$ increases further, the system undergoes a second transition in which the number of zero-energy modes remains equal to two, but their spatial weight shifts back to the edges of the chain. 

\begin{figure}[t!]
    \centering
    \begin{subfigure}{0.23\textwidth}
        \centering
        \includegraphics[width=\textwidth]{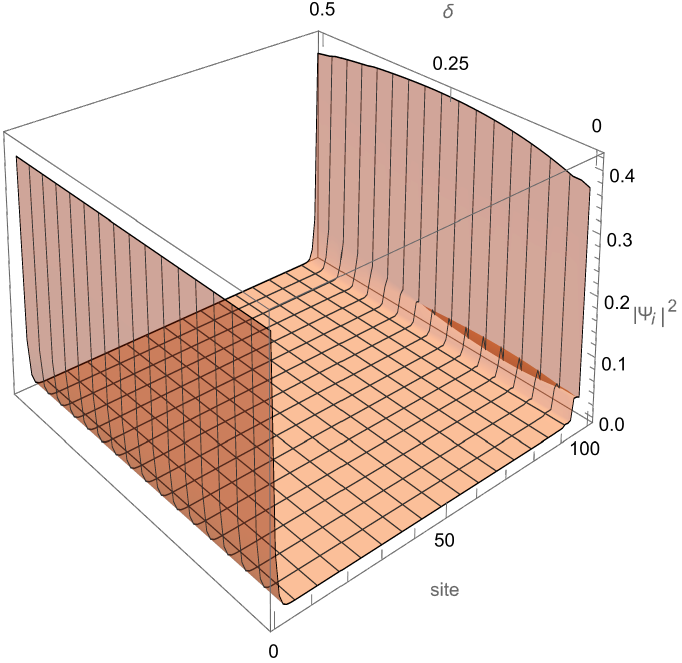}
        \caption*{(a)}
    \end{subfigure}
    \hfill
    \begin{subfigure}{0.23\textwidth}
        \centering
        \includegraphics[width=\textwidth]{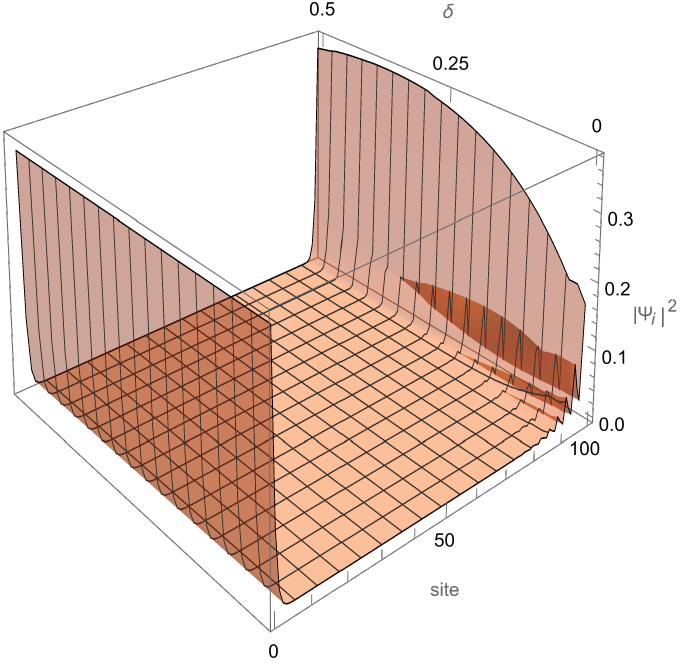}
        \caption*{(b)}
    \end{subfigure}

    \begin{subfigure}{0.23\textwidth}
        \centering
        \includegraphics[width=\textwidth]{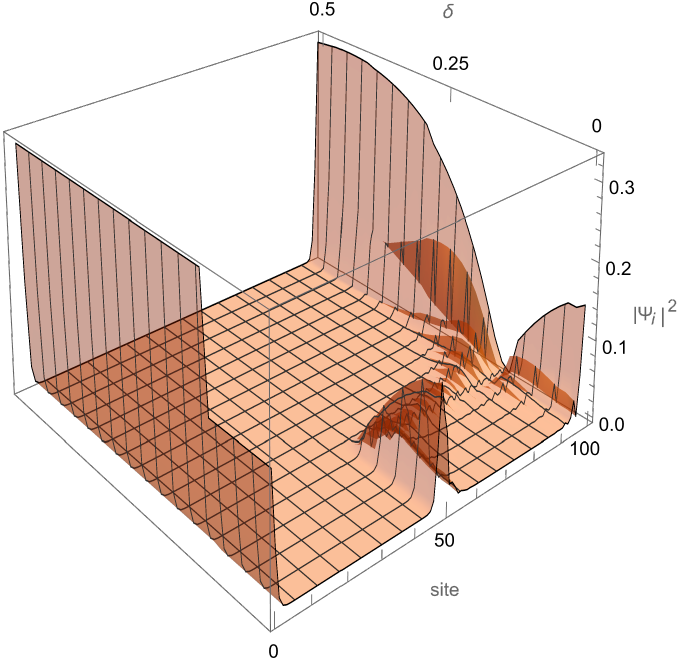}
        \caption*{(c)}
    \end{subfigure}
    \hfill
    \begin{subfigure}{0.23\textwidth}
        \centering
        \includegraphics[width=\textwidth]{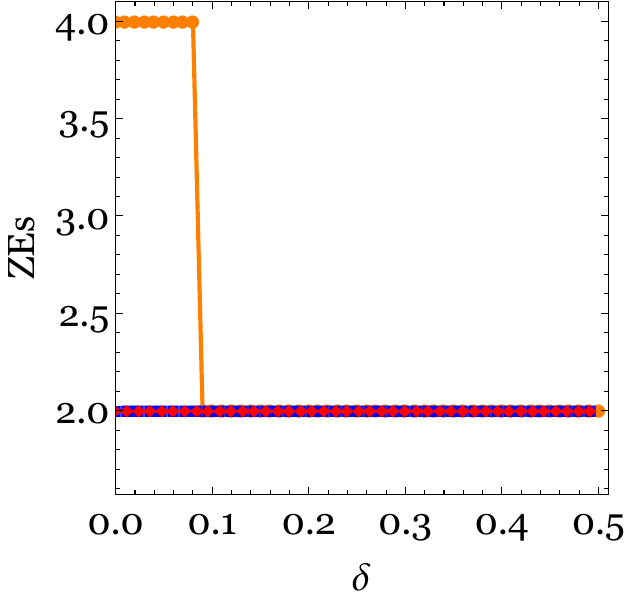}
        \caption*{(d)}
    \end{subfigure}

    \caption{Probability density $\rho_i = |\Psi_i|^2$ at site $i$ as a function of $\delta$ for $\mu = 1$. 
Results are shown for different values of the pairing ratio $\Delta_2/\Delta_1$:  (a) $\Delta_2/\Delta_1 = 0.4$, (b) $\Delta_2/\Delta_1 = 0.6$, (c) $\Delta_2/\Delta_1 = 0.8$, and (d) $\Delta_2/\Delta_1 = 1.2$.}\label{figmu1}
\end{figure}

\paragraph{Generic parameter regime:} Figure~\ref{figmu1} shows the probability densities along the chain for $\mu=1$ as a
function of $\delta$. The same parameter ratios $\Delta_{2}/\Delta_{1}=0.4,\,0.6,\,0.8$
used in Fig.~\ref{figmuzero} are considered here. Figure~\ref{figmu1}(c) shows the emergence of a domain wall at the center of the
chain for $\Delta_{2}/\Delta_{1}=0.8$ and $\mu=1$.
The physics for $\mu\neq 0$ is particularly striking: the combined effect of the
anisotropies and $\delta$ drives a transition from a regime with a single
Majorana zero mode at each edge to one hosting four zero-energy modes
distributed between the edges and the center of the chain. This transition
changes the zero-mode parity sector, taking the system from a single Majorana
pair to two independent Majorana pairs, while leaving the global fermion parity
unchanged.
This mechanism suggests a new way to control the nonlocality of the zero-energy
quasiparticle modes. A slight anisotropy near the optimal value of
$\Delta_{2}/\Delta_{1}$ can, in principle, act as a switch that tunes the degree
of Majorana nonlocality without driving the system into a trivial phase. In
this scenario, the topological character of the state is preserved, while the
nonlocality of the Majorana pair can be continuously modified through the
interplay between the defect and the emergent domain wall. As we discuss below,
this has direct implications for architectures based on topological qubits.

Surprisingly, this physical phenomenon is robust against a finite chemical
potential, which makes our predictions experimentally accessible. 
For the parameter sets used in Figs.~\ref{figmuzero} - \ref{figmu1}, both halves of the chain lie in the same topological phase (with the same integer winding). The domain wall does not separate a trivial and a topological bulk phase. Instead, it realizes an intra-phase solitonic domain wall that redistributes the Majorana zero modes between the edges and the interface, without changing the bulk invariant. Although both halves of the chain share the same bulk topological invariant, the domain wall generates a solitonic mass kink that binds a Majorana mode. Because the bulk topology remains unchanged throughout the process, the Majorana can be displaced between the edge and the interface without closing the gap. This provides a fully local and topologically protected mechanism for Majorana transport within the same phase, which is considerably more robust than schemes relying on global parameter tuning.

\paragraph{Applications for quantum computation:}

Let $\gamma_L$ and $\gamma_R$ denote the Majorana operators localized at the left and right boundaries, respectively. They satisfy $\gamma_{L,R}^\dagger=\gamma_{L,R}$ and $\{\gamma_\alpha,\gamma_\beta\}=2\delta_{\alpha\beta}$.
The Majorana fermionic modes $\gamma_L$ and $\gamma_R$ present zero-energy, undefined charge, and are located at the ends of the chain (one mode at each end - left boundary $\gamma_L$ and right boundary $\gamma_R$ ). These modes combine to form a non-local fermionic mode ($c_i = \frac{1}{2}(\gamma_L+i\gamma_R)$), which is robust against local perturbations or local noises. The non-local fermion
\begin{equation}
f=\frac{\gamma_L+i \gamma_R}{2},\qquad n=f^\dagger f\in\{0,1\},
\label{eq:nonlocal-fermion}
\end{equation}
encodes a qubit in the even/odd \emph{fermion-parity} subspace spanned by $\{\ket{0},\ket{1}\}$ with
\begin{equation}
\hat P = i \gamma_L \gamma_R = 1-2n,\qquad \hat P \ket{0}=+\ket{0},\quad \hat P \ket{1}=-\ket{1}.
\end{equation}
Because $f$ is intrinsically non-local, any \emph{local} perturbation confined to a single edge cannot flip $\hat P$; this is the essence of the topological protection of the encoded information. When a domain wall is introduced, however, the parity $\hat P$ is flipped because the number of zero-energy modes increases and becomes even, with Majorana states now appearing at both edges and at the wall. In this situation the zero mode pairs into an ordinary fermionic quasiparticle, thereby losing its non-local character and the associated topological protection.    
The domain wall provides a practical \emph{control} primitive:
(i) it can \emph{activate} the qubit by ensuring exactly one MZM per edge (odd boundary parity);
(ii) by moving the wall adiabatically (changing $J$) one can \emph{translate} a zero mode along the chain, enabling parity-preserving operations and coupling/decoupling to ancillary elements (e.g., quantum dots) for readout.
In chains  networks (T-junctions), domain walls complement electrostatic gates to implement braiding-like protocols and projective parity measurements without fragile global fine-tuning of $\mu$ or $t_{1,2}$.


\paragraph{Encoding the qubit with Majorana parity:} We formalize the encoding and the ``on/off'' logic enabled by the defect. With open boundaries and within the topological window, the low-energy subspace is spanned by two edge-localized Majoranas $\gamma_{L,R}$. The logical states are
\begin{equation}
\ket{0}_L\otimes\ket{0}_R \equiv \ket{0},\qquad
\ket{1}\equiv f^\dagger\ket{0},\quad f=\frac{\gamma_L+i \gamma_R}{2},
\end{equation}
distinguished by the \emph{global} fermion parity $\hat P=i\gamma_L\gamma_R$.
Any Hamiltonian term strictly local at one edge commutes with $\hat P$ and cannot induce $\ket{0}\!\leftrightarrow\!\ket{1}$ transitions.

\paragraph{Qubit ``on/off'' via domain wall.}
Let $\mathcal{D}$ denote the operation of inserting a single domain wall at cell $J$.
At $\mu=0$ and fixed anisotropy, $\mathcal{D}$ changes the zero-mode accounting by one unit:
it creates a solitonic zero mode $\gamma_J$ near $J$ and toggles the \emph{parity of edge modes}.
When the uniform chain host an \emph{even} number of zero modes per edge (no isolated Majorana), applying $\mathcal{D}$ makes the number \emph{odd}, yielding one Majorana per edge and activating the qubit.
Conversely, removing the wall deactivates it.

\paragraph{Gates and readout (sketch).}
Coupling the two edges through a weak link produces a splitting 
$\varepsilon \propto e^{-L/\xi}$, where $L$ is the chain length and $\xi$ the
coherence length, thereby implementing a $\hat Z$-type rotation via 
$H_{Z}=\tfrac{i}{2}\varepsilon\,\gamma_{L}\gamma_{R}$. A tunable local
coupler near one edge (such as a quantum dot or a superconducting island)
generates $H_{X}\sim i\lambda\,\gamma_{L}\gamma_{J}$ and enables controlled
parity transfer when the domain wall is positioned nearby, with $J$
adiabatically moving along the chain. Readout follows from dispersive shifts or
tunneling spectroscopy that are directly sensitive to the fermion-parity
operator $\hat P$.

\section{Conclusions}
We have identified a fully local and tunable mechanism to control Majorana
qubits in a Kitaev-SSH hybrid chain: a domain wall that redistributes the
zero-mode weight between the edges and the interface without driving the system
out of the topological phase. In the Majorana-supporting regime
$(t_{1,2},\Delta_{1,2})$ and near $\mu \ne0$, this defect acts as a parity
switch that toggles between one and two Majorana pairs, thereby controlling the
degree of nonlocality of the encoded fermionic mode. The solitonic zero-energy
state bound to the wall retains the particle--hole self-conjugacy of Majorana excitation and is protected by the same topological invariant as the edge
modes. These results point to device architectures in which movable domain
walls---rather than electrostatic gating alone---serve as the primary control
primitives for addressing, transporting, and reading out topological qubits. \\


\section*{Acknowledgment}

The authors would like to thank the federal agency FINEP for its financial support through project 3310/2024. This work is also partially supported by the CAPES-PROEX - Finance Code 001, CNPq and FAPEAM agencies.

\bibliographystyle{IEEEtran}
\bibliography{bibitex}

\end{document}